\documentstyle[epsfig,12pt,aaspp4]{article}
\lefthead{Naik et al.}
\righthead{$Suzaku$ observation of A0535+262}
\slugcomment{}

\received{}
\begin{document}

\title{Broad band X-ray spectroscopy of A0535+262 with $Suzaku$}
\author{S. \textsc{Naik}\altaffilmark{1},
T. \textsc{Dotani}\altaffilmark{1},
Y. \textsc{Terada}\altaffilmark{2},
M. \textsc{Nakajima}\altaffilmark{2},
T. \textsc{Mihara}\altaffilmark{2},
M. \textsc{Suzuki}\altaffilmark{2},
K. \textsc{Makishima}\altaffilmark{2,3},
K. \textsc{Sudoh}\altaffilmark{4},
S. \textsc{Kitamoto}\altaffilmark{4},
F. \textsc{Nagase}\altaffilmark{1},
T. \textsc{Enoto}\altaffilmark{3},
H. \textsc{Takahashi}\altaffilmark{5}
}
\altaffiltext{1}{Institute of Space and Astronautical Science, JAXA, 
     3-1-1 Yoshinodai, Sagamihara, Kanagawa 229-8510, 
     JAPAN. email : naik@astro.isas.jaxa.jp}
\altaffiltext{2}{Cosmic Radiation Laboratory, Institute of Physical and 
    Chemical Research, Wako, Saitama 351-0198, Japan}
\altaffiltext{3}{Department of Physics, University of Tokyo, 7-3-1 Hongo, Bunkyo-ku, Tokyo 113-0033, Japan}
\altaffiltext{4}{Department of Physics, Rikkyo University, 3-34-1, 
Nishi-Ikebukuro, Toshima-ku, Tokyo 171-8501, Japan}
\altaffiltext{5}{Department of Physical Science, Hiroshima University, 
1-3-1 Kagamiyama, Higashi-Hiroshima, Hiroshima 739-8526, Japan}

\begin{abstract}
The transient X-ray binary pulsar A0535+262 was observed with $Suzaku$ on
2005 September 14 when the source was in the declining phase of the
August-September minor outburst. The $\sim$103 s X-ray pulse profile was 
strongly energy dependent, a double peaked profile at soft X-ray energy 
band ($<$ 3 keV) and a single peaked smooth profile at hard X-rays. The
width of the primary dip is found to be increasing with 
energy. The broad-band energy spectrum of the pulsar is well described 
with a Negative and Positive power-law with EXponential (NPEX) continuum 
model along with a blackbody component for soft excess. A weak iron 
K$_\alpha$ emission line with an equivalent width $\sim$ 25 eV was detected 
in the source spectrum. The blackbody component is found to be pulsating  
over the pulse phase implying the accretion column and/or the inner edge 
of the accretion disk may be the possible emission site of the 
soft excess in A0535+262. The higher value of the column density is 
believed to be the cause of the secondary dip at the soft X-ray energy band. 
The iron line equivalent width is found to be constant (within errors) over 
the pulse phase. However, a sinusoidal type of flux variation of iron emission 
line, in phase with the hard X-ray flux suggests that the inner accretion 
disk is the possible emission region of the iron fluorescence line. 
\end{abstract}


\section{Introduction}
Be X-ray binaries consist of a neutron star in an eccentric orbit
around a Be star companion. The orbit of the Be X-ray binaries is generally
wide and eccentric with orbital periods in the range of 16 days to 400 days 
(Coe 2000). Mass transfer from the Be companion to the neutron star takes 
place through the circumstellar disk. When the neutron star passes through 
the disk or during the periastron passage, it shows strong outbursts with 
an increase in X-ray luminosity by a factor $\geq$ 100 (Negueruela 1998).

A0535+262 is a 103 s Be/X-ray binary pulsar discovered by $Ariel~V$ during a
large outburst in 1975 (Coe et al. 1975). The binary companion HDE~245770 is 
an O9.7-B0 IIIe star in a relatively wide eccentric orbit ($e$ = 0.47) with 
orbital period of $\sim$ 111 days and at a distance of $\sim$ 2 kpc (Finger 
et al. 1996; Steele et al. 1998). The pulsar shows regular outbursts with the 
orbital periodicity. Occasional giant X-ray outbursts are also observed when 
the object becomes even brighter than the Crab. The pulsar shows three 
typical intensity states, such as quiescence with flux level of below 10 mCrab,
normal outbursts with flux level in the range 10 mCrab to 1 Crab, and giant 
outbursts during which the object becomes the brightest X-ray source in the 
sky with the flux level of several Crab (Kendziorra et al. 1994). 

103 s pulsations were detected during past X-ray observations of A0535+262
in quiescence, outburst, and giant outbursts. The pulse profile  
was single peaked in quiescence (in 1-10 keV range; Mukherjee \& Paul 2005, 
3-20 keV; Negueruela et al. 2000), double peaked with a clearly asymmetric
''main'' and a more symmetric ''secondary'' pulse during the X-ray outbursts
(Mihara 1995; Kretschmar et al. 1996, Maisack et al. 1997). The X-ray spectrum 
of the pulsar has been studied at soft and hard X-rays with various instruments
at different luminosity levels. The spectrum of the object, during outbursts, 
shows cyclotron resonant scattering features at higher energies than that of
the other pulsars. Two harmonic features at around 50 keV and 100 
keV were detected in its 1989 outburst with the HEXE/TTM instrument on 
Mir/Kvant (Kendziorra et al. 1994). The CGRO/OSSE observations of 1994
outburst of the pulsar showed a significant absorption feature at 
110 keV (Grove et al. 1995). These detections did not resolve whether 
the magnetic field of the pulsar is $\sim$ 5 $\times$ 10$^{12}$ G (when 
the fundamental occurs at 55 keV) or $\sim$10$^{13}$ G (for the 110 keV 
fundamental). 

The most recent major X-ray outburst was detected in 2005 May/June with the 
BAT instrument on Swift when the 15-195 keV count rate was greater than 3 
times that of the Crab Nebula (Tueller et al. 2005). Following the detection
of the recent outburst, the pulsar was observed by INTEGRAL, RXTE and
the recently launched $Suzaku$. A cyclotron resonance feature at $\sim$45 
keV was detected in the Hard X-ray Detector (HXD) spectrum of $Suzaku$
with estimated magnetic field of the pulsar as $\sim$ 4 $\times$ 10$^{12}$
Gauss (Terada et al. 2006). The detection of the absorption feature at 
$\sim$45 keV and its first harmonic at $\sim$100 keV is reported from 
the INTEGRAL and RXTE observations of the pulsar during the 2005 
August/September outburst (Caballero et al. 2007). Using the same Suzaku
observation used for the analysis of the cyclotron resonance feature
(Terada et al. 2006), we study the broad-band spectral properties of 
the pulsar in the present paper.

\section{Observation}
\begin{figure}
\vskip 7.9 cm
\includegraphics{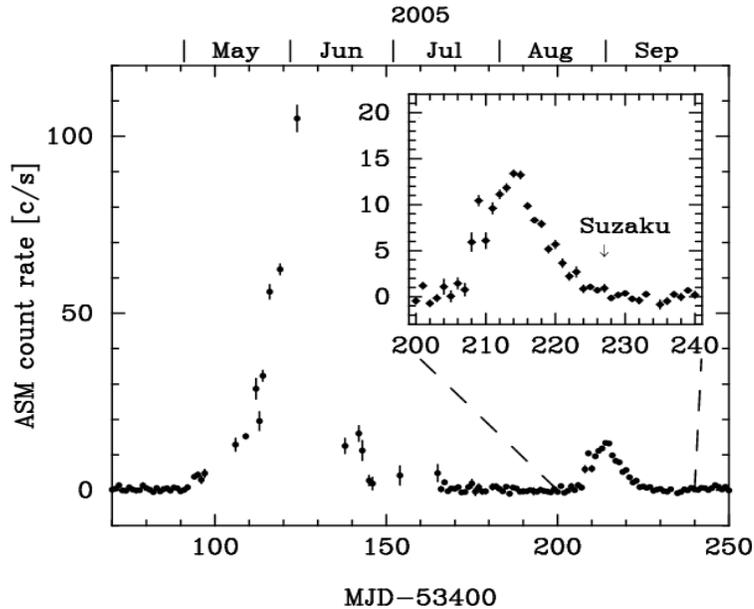}
\caption{The RXTE-ASM one-day averaged light curve of A0535+262 from 2005 
April 10 (MJD 53470) to 2005 October 07 (MJD 53650). The inset shows the 
expanded version of the minor outburst in 2005 August-September with the arrow 
showing the time of the Suzaku observation of the pulsar.}
\label{asm}
\end{figure}

The detection of the recent outburst of A0535+262 on 16 May 2005 with the 
BAT instrument on Swift prompted many observatories to observe the pulsar 
during this period. The RXTE/ASM monitoring of the pulsar showed one major
outburst of the pulsar which lasted from 2005 May 06 to June 24. During this
outburst, the peak luminosity was about 1.4 Crab (103 ASM counts s$^{-1}$).
Approximately after one orbital period of 111 days, the pulsar was again 
detected with the RXTE/ASM and RXTE/PCA with peak luminosity of about 0.2 
Crab (12 ASM counts s$^{-1}$). During this minor outburst, the pulsar was 
observed with the RXTE, INTEGRAL, and $Suzaku$. The RXTE/ASM one-day 
averaged light curve of A0535+262 between 2005 April 10 and 2005 October 07 
is shown in Figure~\ref{asm}. The arrow mark in the inset figure indicate 
the observation of the pulsar with $Suzaku$. During the declining phase of 
the minor outburst, A0535+262 was observed with $Suzaku$ on 2005 September 
14 from 13:40 UT to 01:00 UT on the next day (orbital phase range of 
0.42-0.43; Finger et al. 1994). This Target of 
Opportunity (TOO) observation was carried out at ``XIS nominal'' pointing 
position for effective exposures of 22.3 ksec with the XIS and 21.7 ksec with 
the HXD. The XIS was operated with ``1/4 window'' option which gives a time 
resolution of 2 sec, covering a field of view of 17$'$.8$\times$4$'$.4.

$Suzaku$, the fifth Japanese X-ray astronomy satellite (Mitsuda et al. 2007), 
was launched on 2005 July 10. It covers 0.2--600 keV energy range with the two 
sets of instruments, X-ray Imaging Spectrometer (XIS; Koyama et al. 2007) 
covering the soft X-rays in 0.2-12 keV energy range, and the Hard X-ray 
Detector (HXD; Takahashi et al. 2007) which covers 10--70 keV with PIN 
diodes and 30--600 keV with GSO scintillators. There are 4 sets of XIS, each 
with a 1024 $\times$ 1024 pixel X-ray-sensitive CCD detector at the focus of  
each of the four X-ray Telescopes (XRT). One of the four CCDs is back 
illuminated (BI) whereas the other three are front illuminated (FI). The 
field of view of the XIS is 18'$\times$18' in a full window mode 
with an effective area of 340 cm$^2$ (FI) and 390 cm$^2$ (BI) at 1.5 keV. 
The energy resolution was 130 eV (FWHM) at 6 keV just after the launch. 
The HXD is a non-imaging instrument that is designed to detect high-energy 
X-rays. The HXD has 16 identical units made up of two types of detectors, 
silicon PIN diodes ($<$ 70 keV) and GSO crystal scintillator ($>$ 30 keV). 
The effective areas of PIN and GSO detectors are $\sim$ 145 cm$^2$ at 15 keV 
and 315 cm$^2$ at 100 keV respectively. For a detailed description of the XIS 
and HXD detectors, refer to Koyama et al. (2007) and Takahashi et al. (2007).

\section{Analysis and Results}

The HXD data of the $Suzaku$ observation of A0535+262 has already been 
presented by Terada et al. (2006). The same procedure was followed for 
the HXD/GSO data reduction to obtain the GSO source and background spectra. 
For HXD/PIN and XIS data reduction, we used the cleaned event data (ver 
1.2 products) to obtain the PIN and XIS light curves and source spectra. 
The simulated background events (bgd$_{-}$a) were used to estimate the 
HXD/PIN background (Kokubun et al. 2007) for the A0535+262 observation. 
The response files, released in March 2006 and August 2006 were used for 
HXD/GSO, and HXD/PIN spectra, respectively. The accumulated events of the 
XIS data were discarded when the telemetry was saturated, data rate was 
low, the satellite was in the South Atlantic Anomaly (SAA), and when the 
source elevation above the earth's limb was below 5$^\circ$ for night-earth 
and below 20$^\circ$ for day-earth. Applying these conditions, the source 
spectra were accumulated from the XIS cleaned event data by selecting a 
circular region of 4$'$.3 around the image centroid. Because this extraction 
circle is larger than the optional window, the effective extraction region 
is the intersection of the window and this circle. The XIS background spectra 
were accumulated from the same observation by selecting rectangular regions 
away from the source. The response files and effective area files for XIS 
were generated by using the ''xissimarfgen'' and ''xisrmfgen'' task of 
FTOOLS (V6.2). X-ray light curves of 2 s time resolution were extracted 
from the XIS and PIN event data. In the ver 1.2 products of Suzaku observations,
it is known that the XIS time assignment contains an error of 6 seconds compared
to HXD when the '1/4 window' option was applied\footnote{http://www.astro.isas.jaxa.jp/suzaku/analysis/xis/timing/}. The HXD absolute time assignment was
verified to be correct better than 360 $\mu$s with the observation of the Crab
pulsar (Terada et al. 2007). Although the time assignment error in XIS 
was much smaller than the pulse period of A0535+262, we corrected it before 
the analysis of the XIS data.

\begin{figure}
\vskip 8.0 cm
\includegraphics{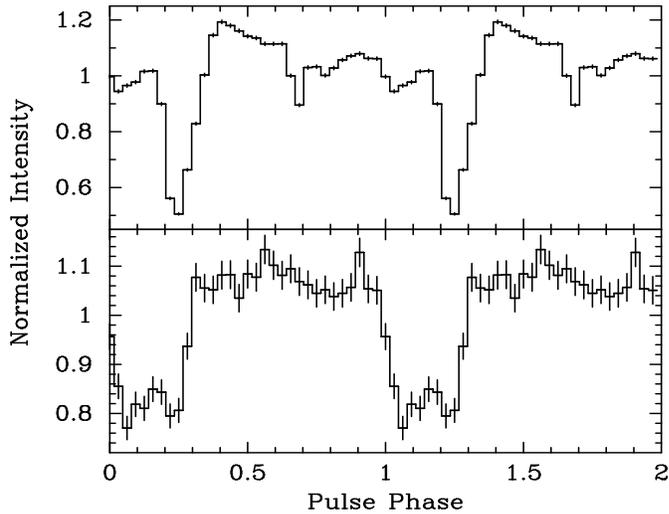}
\caption{The XIS (upper panel) and PIN (lower panel) pulse profiles of 
A0535+262 from the $Suzaku$ observation of the minor outburst in 2006 
August-September in 0.2--12 keV and 8-70 keV respectively. The light curves 
are not background subtracted. The error bars represent 1 sigma uncertainties.
Two pulses in each panel are shown for clarity.}
\label{pp}
\end{figure}

\subsection{Timing Analysis}

\begin{figure*}
\vskip 12.0 cm
\includegraphics{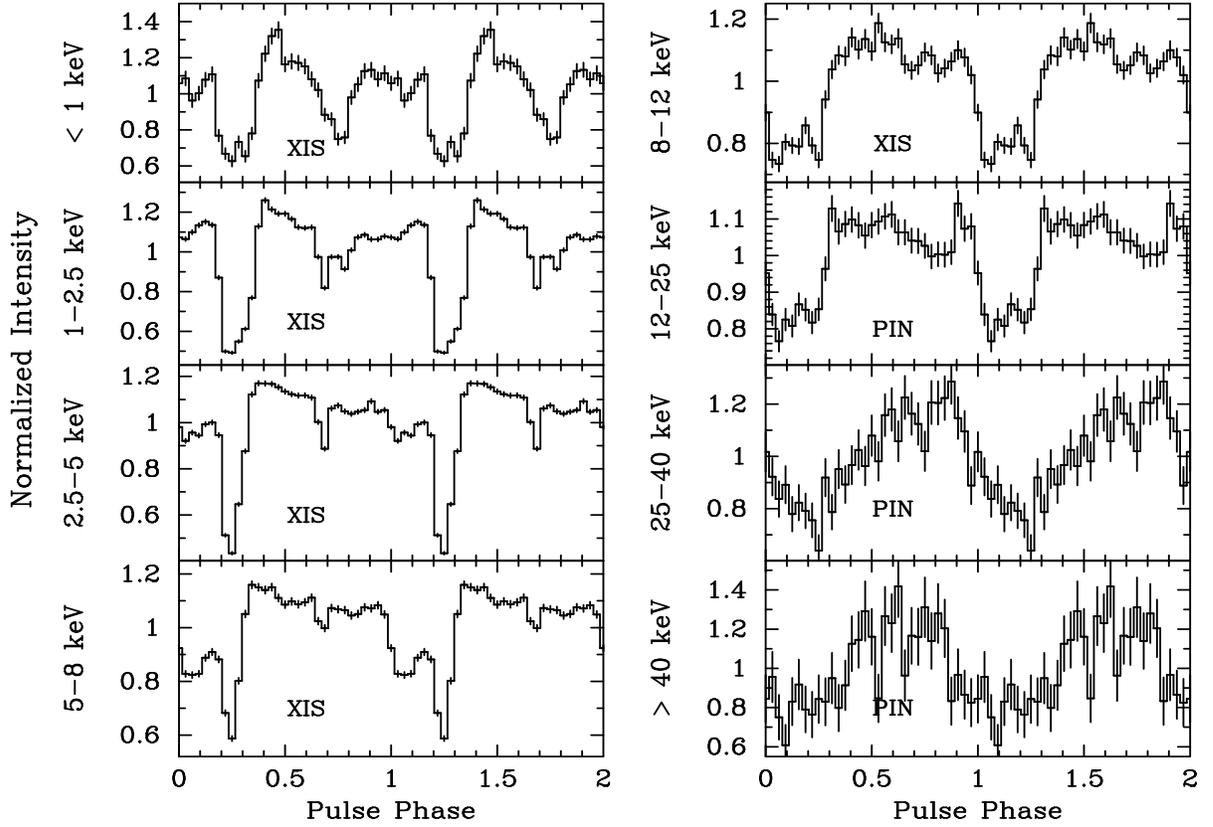}
\caption{The XIS and HXD/PIN pulse profiles of A0535+262 at different
energy bands. No background was subtracted from the folded profiles. We can
see the presence/absence of the dip like structure in 0.65-0.80 pulse phase 
range. Apart from this structure, a gradual decrease in the normalized 
intensity in 0.0-0.17 phase range and finally merging with the primary dip 
in 0.17-0.35 phase range (primary dip) can be seen. The error bars 
represent 1 sigma uncertainties. Two pulses in each panel are shown for 
clarity.}
\label{erpp}
\end{figure*}

Barycentric correction was applied to the light curves of A0535+262 for the 
measurement of the pulse period. Pulse folding and a $\chi^2$ maximization
method was applied to the light curves yielding the pulse period of the
pulsar to be 103.375$\pm$0.09 s (as reported in Terada et al. 2006). The pulse 
profiles obtained from the XIS and PIN light curves of the $Suzaku$ 
observation of the pulsar are shown in Figure~\ref{pp}. From the figure, 
it is observed that the shape of the pulse profile in the XIS energy band 
(0.2 -- 12 keV) is different from that in the HXD/PIN energy band (8-70 keV). 
The dip in the pulse profile in pulse phase range 0.17-0.35, hereafter 
referred to as a primary dip, is narrow at soft X-rays (XIS energy band) 
and broad in the hard X-ray energy band. Apart from the variable width 
of the dip, a dip like structure is present in the soft X-ray pulse profile 
in 0.65-0.80 pulse phase range which is absent in the hard X-ray profile. 
To investigate the energy dependence of the pulse profile of A0535+262, we 
generated light curves in different energy bands from XIS and PIN event data. 
The light curves are folded with the pulse period and the corresponding 
pulse profiles are shown in Figure~\ref{erpp}. The energy resolved pulse 
profiles of the pulsar obtained from $Suzaku$ observation are found to be 
different from that of the previous observations (Mukherjee \& Paul 2005; 
Negueruela et al. 2000; Kretschmar et al. 1996; Mihara 1995). The dip like 
structure (in pulse phase range 0.65-0.80) is found to be very prominent 
below 1 keV. The width and depth of this structure decrease gradually with 
energy up to $\sim$8 keV, beyond which it becomes indistinguishable from the
fine structures in the pulse profile. A gradual decrease in the normalized 
intensity in 0.0-0.17 pulse phase range (as shown in the figure) is found 
at soft X-ray profiles which finally merged with the primary dip, making 
the hard X-ray pulse profiles smooth and single peaked. Pulse phase resolved 
spectroscopy would help in understanding the nature of the dip like structure
and the gradual decrease of normalized intensity with energy prior to the 
primary dip in the pulse profile of A0535+262.

\subsection{Spectral Analysis}

\subsubsection{Pulse phase averaged spectroscopy}
We calculated the energy spectrum from the events
selected in the energy ranges of 0.3-10.0 keV for the back 
illuminated CCD (XIS-1), 0.5-10.0 keV for front illuminated CCDs 
(XIS-0, 2, 3), 10-70 keV for the HXD/PIN and 45-135 keV for the HXD/GSO
detectors. After appropriate background subtraction, simultaneous spectral 
fitting was done using the XIS, PIN and GSO spectra. All the spectral 
parameters other than the relative normalization, were tied together for 
all the detectors. We tried to fit the broad-band energy spectra with a model 
consisting of a blackbody component, a power-law with exponential cutoff, 
a Gaussian function for the iron fluorescence line at 6.4 keV and the 
cyclotron resonance factor (CYAB) at $\sim$45 keV (as detected by Caballero 
et al. 2007). The analytical form of CYAB is

\begin{equation}
CYAB(E) = \exp{ \biggl\{ - \frac{\tau~(WE/E_a)^2}{(E-E_a)^2+W^2} \biggl\} },
\label{eq2}
\end{equation}

where $E_a$ is the resonance energy, $W$ is the width of
the absorption structure, and $\tau$ is the depth of the resonance.

This model gave a reduced $\chi^2$ of 1.41 (for 804 dof). The parameters of 
the cyclotron resonance feature are found to be consistent with that 
reported in Terada et al. (2006). It is found that the residual structures
below 1 keV contribute significantly to the statistics of the spectral
fitting. Because the spectral shape in this energy range is determined mainly
by the low-energy absorption, detailed modeling of the absorber (in terms of
abundance, uniformity, etc.) may be necessary to improve fitting. However,
we are interested in the global shape of the energy spectrum, not in the local 
structures, so we simply ignored the data below 1 keV for the subsequent 
spectral fitting. Simultaneous spectral fitting to the XIS and HXD spectra 
in 1.0-135.0 keV with above model improved the fitting result with a 
reduced $\chi^2$ of 1.33 (for 733 dof). However, a blackbody temperature of 
$\sim$1.36 keV is unusual in accretion-powered X-ray pulsars. This does not
directly mean that the model is unacceptable, but it is worth pursuing another
model. We have also tried to fit the continuum spectrum using a power-law 
with a Fermi-Dirac cutoff, as described by Coburn et al. (2002) 
along with a blackbody component, Gaussian function, and cyclotron resonance 
factor. The spectral fitting, however, was very poor with reduced $\chi^2$ of 
2.2 (for 736 dof).

We then tried to fit the broad band spectrum of A0535+262 using the NPEX 
continuum component along with the cyclotron resonance factor (CYAB) and 
interstellar absorption. The analytical forms of the NPEX model is 
\begin{equation}
NPEX(E) = (N_1 E^{-\alpha_1} + N_2 E^{+\alpha_2})  ~ \exp \left( -\frac{E}{kT} \right),
\label{eq1}
\end{equation}
where $E$ is the X-ray energy (in keV), $N_1$ and $\alpha_1$ are the
normalization and photon index of the negative power-law respectively, 
$N_2$ and $\alpha_2$ are those of the positive power-law, and $kT$ is the
cutoff energy in units of keV. All five parameters of the NPEX continuum 
component were kept free. However, as in the case of Terada et al. (2006), 
we could not constrain the positive power-law index ($\alpha_2$) very well. 
Terada et al. fixed $\alpha_2$ to 2, but we found that a little larger 
value of $\alpha_2$ improved the overall fit to the data. The value 
of $\chi^2$ found by fixing the value of $\alpha_2$ to 2 and 3 are 
1029 and 933 for 730 degrees of freedom respectively. Therefore, we 
fixed the second index ($\alpha_2$) to 3 in the subsequent analysis.
The relative instrument normalizations of the four XISs, PIN and GSO 
detectors were kept free and the values are found to be  
1.0:1.0:0.92:1.0:1.1:1.04 for XIS3:XIS0:XIS1:XIS2:PIN:GSO with a
clear agreement with the detector calibration. 

\begin{figure*}
\vskip 12.0 cm
\includegraphics{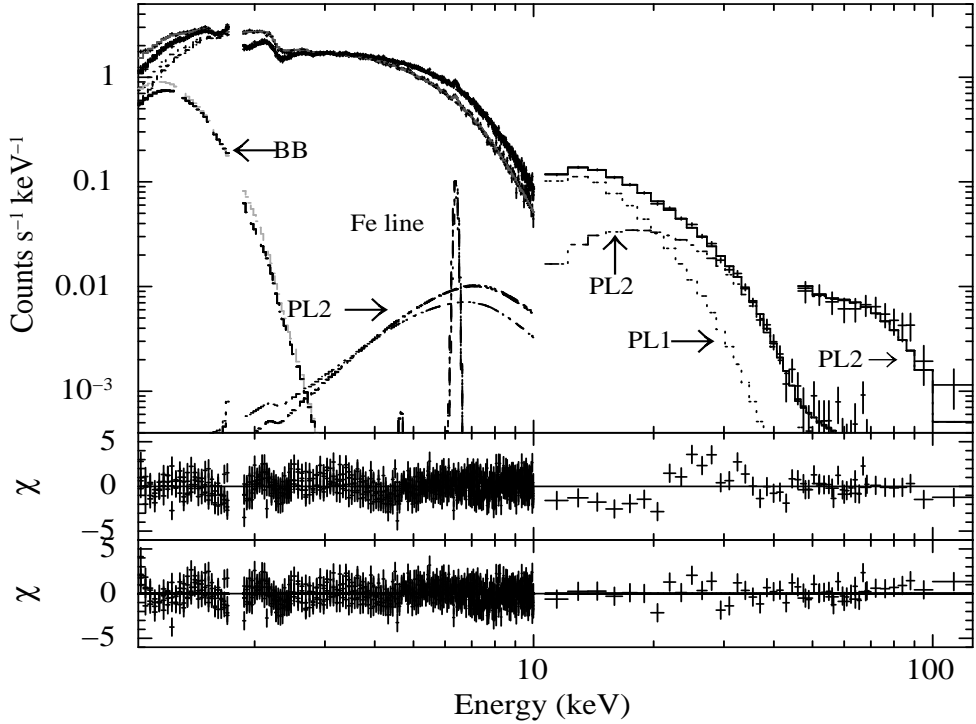}
\caption{Energy spectrum of A0535+262 obtained with the XIS, PIN, and GSO
detectors of the $Suzaku$ observation, along with the best-fit model
comprising a blackbody component (BB), Negative Positive Exponential (NPEX)
power law continuum model and a narrow iron line emission (Fe line) and a 
cyclotron resonance factor. The negative and positive power laws are marked
by PL1 and PL2 respectively. The middle and bottom panels show the 
contributions of the residuals to the $\chi^2$ for each energy bin for the
cutoff power-law continuum model and the NPEX continuum model respectively.}
\label{spec}
\end{figure*}

The CYAB parameters were initially set around the value quoted in 
Terada et al. (2006) and then allowed to find the corresponding best-fit 
values. The NPEX continuum model showed significant improvement over 
the Fermi-Dirac cutoff and cutoff power-law models in the spectral 
fitting with reduced $\chi^2$ of 1.28 for 730 dof. It was found that 
the NPEX continuum model fits the Suzaku 1.0-135 keV spectrum better than 
the cutoff power-law continuum model which was used to described the RXTE 
and INTEGRAL spectra of the pulsar (Caballero et al. 2007). The spectral 
parameters of the best-fit model obtained from the simultaneous spectral 
fitting to the XIS, PIN, and GSO data of the $Suzaku$ observation of 
A0535+262 are given in Table~\ref{spec_par}. The count rate spectra of the 
$Suzaku$ observation is shown in Figure~\ref{spec} along with the model 
components (top panel) and residuals to the cutoff power-law continuum model
(middle panel) and the best-fit NPEX continuum model (bottom panel). We need 
to note that the iron line width could be an artifact, because no degradation 
of the energy resolution is included in the response files. In such a case, 
an artificial line width of ~20 eV (1$\sigma$, at 5.9 keV) may be obtained 
even at the time of A0535+262 observation (Koyama et al. 2007).

\begin{table}
\caption{Best-fit parameters of the phase-averaged spectra for A0535+262 
during 2005 $Suzaku$ observation with 1$\sigma$ errors}
\begin{tabular}{ll}
\hline
\hline
Parameter	&Value\\
\hline
\hline
N$_H$ (10$^{22}$ atoms cm$^{-2}$)    &1.10$\pm$0.01\\
$kT_{BB}$ (keV)                      &0.16$\pm$0.01\\
Iron line Energy (keV)           &6.39$\pm$0.01\\
Iron line width  (keV)           &0.03$\pm$0.02\\
Iron line eq. width (eV)         &25$\pm$3\\
Iron line flux$^a$               &0.8$\pm$0.1\\
Power-law index ($\alpha_1$)     &0.84$\pm$0.02\\
Power-law normalization (N$_1$)	 &3.21$_{-0.02}^{+0.05}$ $\times$ 10$^{-2}$\\
Power-law normalization (N$_2$)	 &3.1$\pm$0.2 $\times$ 10$^{-7}$\\
Exponential cutoff energy (keV)      &9.12$\pm$0.06\\
CYAB energy ($E_a$, in keV)          &44.8$_{-0.7}^{+1.3}$\\
CYAB depth  ($\tau$)                 &1.6$_{-0.13}^{+0.26}$\\
CYAB width ($W$, keV)                &9.5$\pm$1.2\\
Blackbody flux$^a$                   &3.4$\pm$0.2\\
Source flux$^b$                      &9.4$_{-0.3}^{+0.4}$\\
Reduced $\chi^2$		     &1.28 (730 dof)\\
\hline
\end{tabular}
\\
$N_1$ and $N_2$ are the normalizations of negative and positive power-laws respectively with $\alpha_2$ kept fixed at 3\\
$^a$ : in 10$^{-12}$  ergs cm$^{-2}$ s$^{-1}$ unit and estimated in 1.0-10.0 keV energy range.\\
$^b$ : in 10$^{-10}$  ergs cm$^{-2}$ s$^{-1}$ unit and estimated in 1.0-70.0 keV energy range. \\
\label{spec_par}
\end{table}

\subsubsection{Pulse phase resolved spectroscopy}

The presence of significant energy dependent dips in the pulse profile of 
A0535+262 prompted us to make a detailed study of the spectral properties 
at different pulse phases of the pulsar. To investigate the changes
in the spectral parameters at soft X-rays at different pulse phases of the 
pulsar, we used data from the XIS (both BI and FIs) and HXD/PIN detectors.
The XIS and PIN spectra were accumulated into 10 pulse phase bins by
applying phase filtering in the FTOOLS task XSELECT. The XIS and PIN 
background spectra and response matrices used for the phase averaged 
spectroscopy, were also used for the phase resolved spectroscopy.
Simultaneous spectral fitting was done in the 1.0-70.0 keV energy band.

\begin{figure*}
\vskip 12.0 cm
\includegraphics{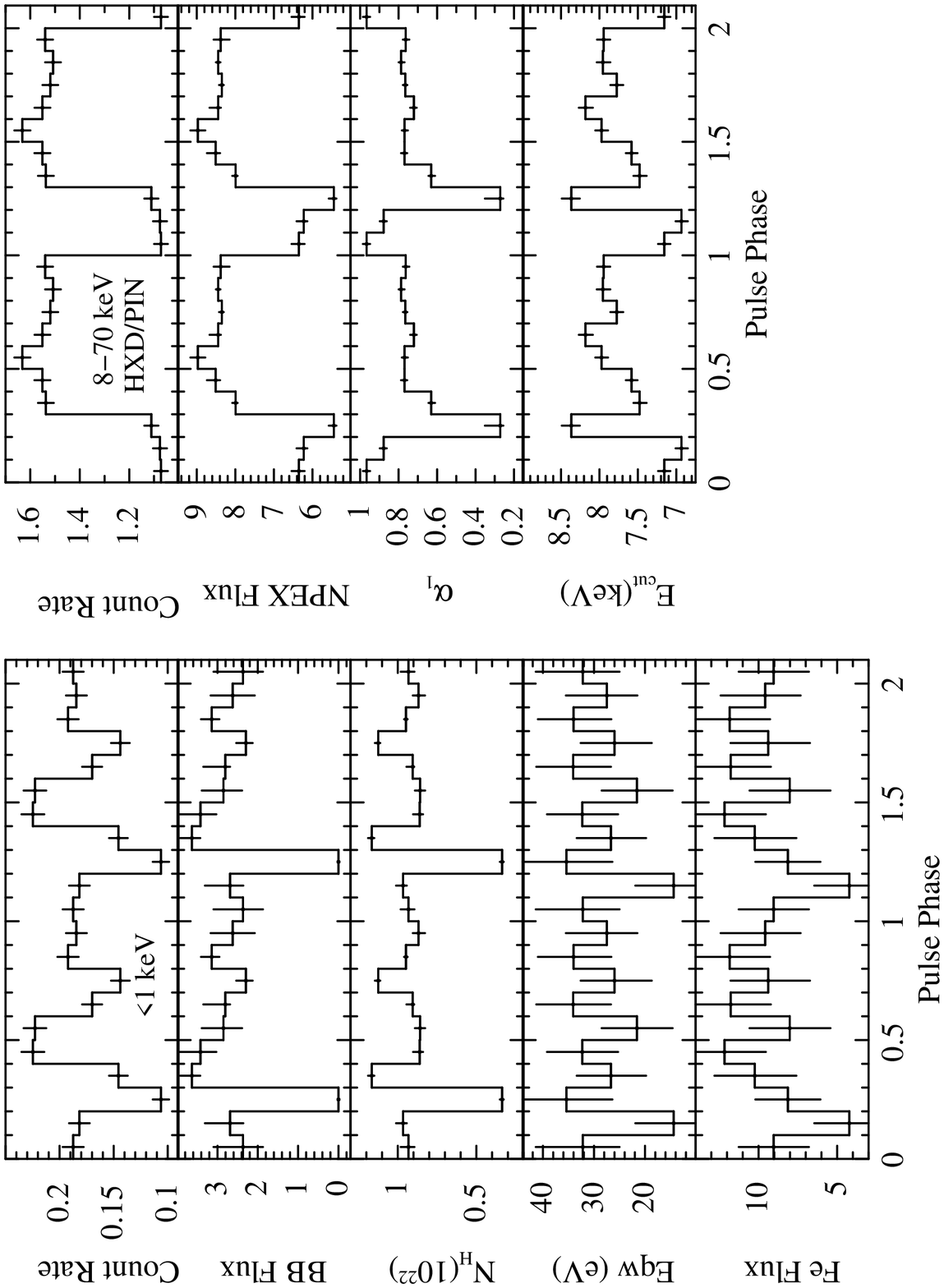}
\caption{Spectral parameters obtained from the pulse phase resolved
spectroscopy of $Suzaku$ observation of A0535+262. The errors shown 
in the figure are estimated for 1$\sigma$ confidence level. In the 
figure, the blackbody flux (BB Flux) in 1-10 keV energy band and iron 
line flux (Fe Flux) are plotted in the units of 10$^{-12}$ and 10$^{-13}$ 
ergs cm$^{-2}$ s$^{-1}$ respectively, whereas the NPEX flux in 1-50 keV 
energy band is plotted in the units of 10$^{-10}$ ergs cm$^{-2}$ 
s$^{-1}$. The blackbody flux and the NPEX flux are not corrected for 
low-energy absorption and cyclotron feature at $\sim$45 keV. The XIS 
($<1$ keV) and PIN (8-70 keV) count rates are also shown in the left 
top and right top panels respectively.}
\label{phrs}
\end{figure*}

The phase resolved spectra were fitted with the same continuum model used to 
describe the phase averaged spectrum. Because of the coupling among the 
parameters, we had to fix some of them to get meaningful constraints to the 
model parameters. The CYAB model parameters, the blackbody temperature, the 
iron line energy and line width were kept fixed at the phase averaged values. 
The positive power law index ($\alpha_2$) was fixed at 3, as in the case of
the phase-averaged spectroscopy. The ratio of the negative and positive 
power-law normalizations (N$_1$/N$_2$) was also fixed at the value obtained 
from the phase averaged spectroscopy. The relative instrument normalizations 
were fixed at the values obtained from the phase averaged spectroscopy. The 
parameters obtained from the simultaneous spectral fitting to the XIS and PIN 
phase resolved spectra are shown in Figure~\ref{phrs} along with the XIS 
($<1$ keV) and PIN (8-70 keV) count rates at left top and right top panels. 
We note that the value of $N_H$ in phase 0.2-0.3 ($\sim$3$\times$10$^{21}$
atoms cm$^{-2}$) is lower than the galactic hydrogen column density towards
the pulsar ($\sim$5.9$\times$10$^{21}$ atoms cm$^{-2}$). However, because
the interstellar medium can be highly non-uniform, we consider that this
$N_H$ value is within the acceptable range.

\begin{figure}[h]
\vskip 7 cm
\includegraphics{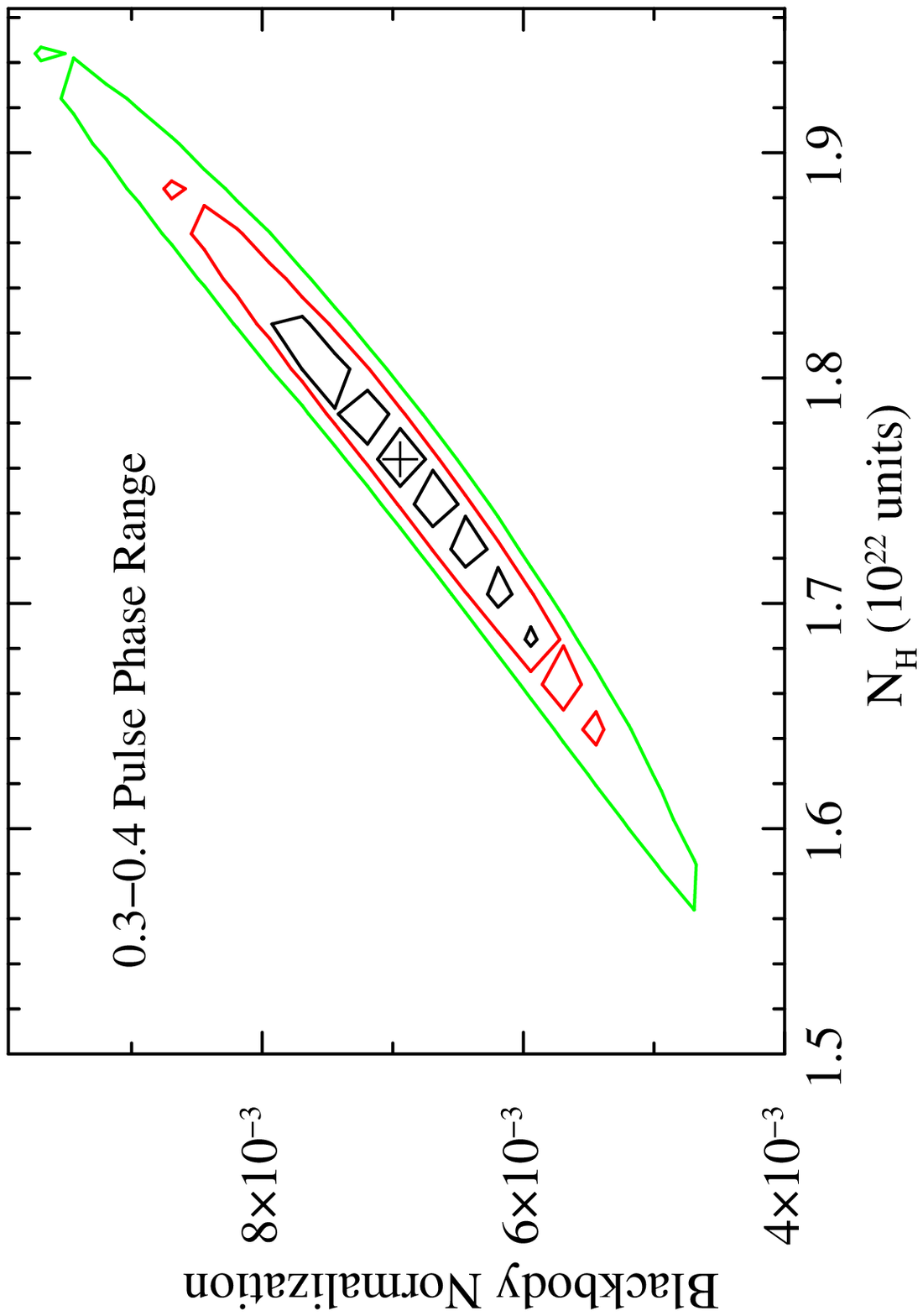}
\includegraphics{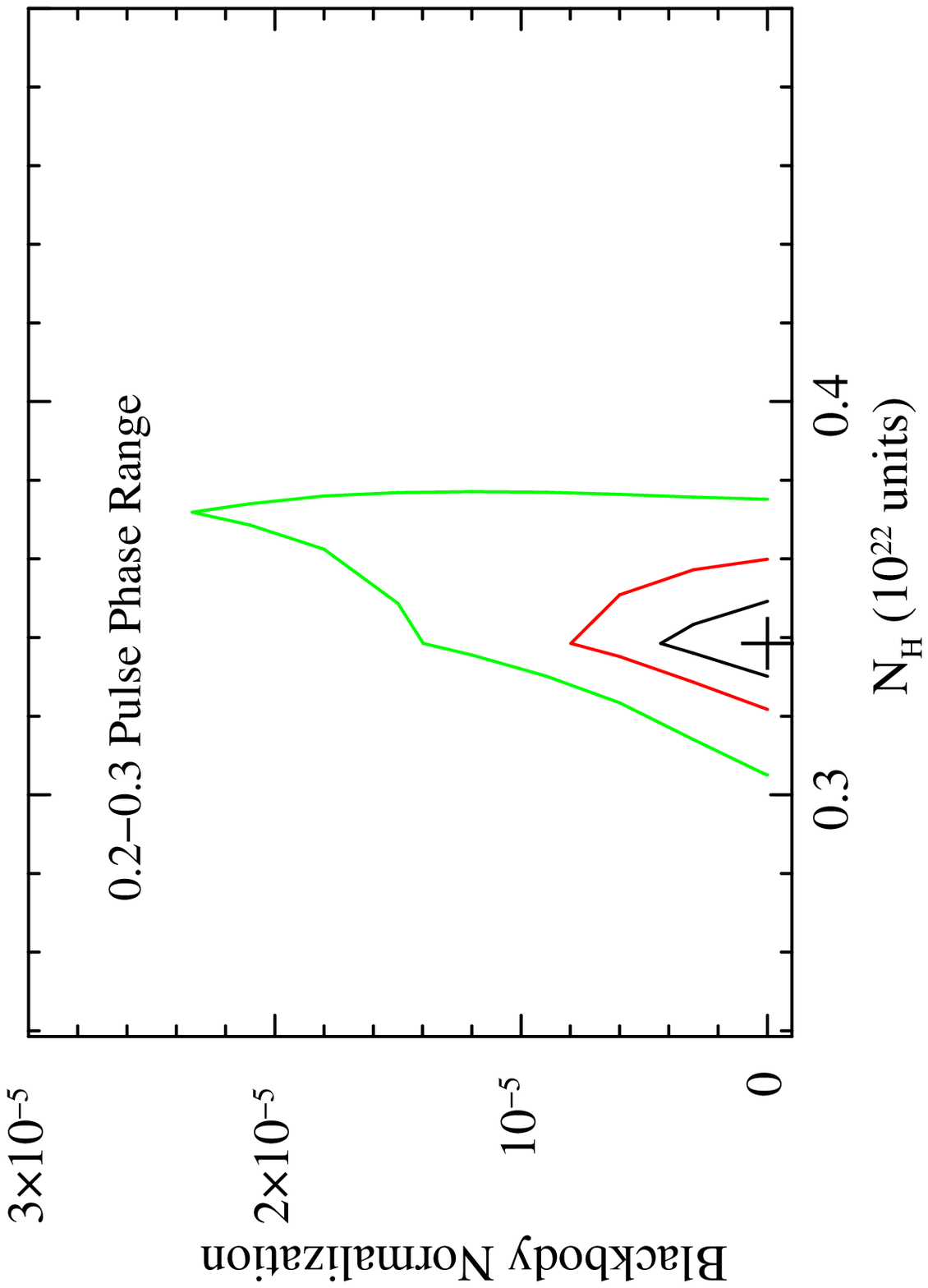}
\caption{Blackbody normalization vs. the absorption column density (N$_H$),
obtained using spectral fits of the phase resolved spectra in 0.3-0.4 (left 
panel) and 0.2-0.3 (right panel) pulse phase ranges. The blackbody normalization
is defined as L$_{39}$/D$_{10}^2$, where L$_{39}$ is the source luminosity 
in units of 10$^{39}$ erg/s, and D$_{10}$ is the distance to the source in 
the unit of 10 kpc. Three contours correspond to 68\%, 90\%, and 99\% 
confidence limits for two interested parameters.}
\label{cont}
\end{figure}

Minimum values for the blackbody flux, absorption column density, and photon 
index $\alpha_1$ occur at phase 0.2-0.3. Although the minimum looks 
significant, we should be careful about the correlation among the parameters.
In Figure~\ref{cont}, we plot confidence contours between N$_H$ and the 
blackbody normalization (N$_{BB}$) to check the acceptable ranges of these 
parameters taking two extreme phases, namely 0.3-0.4 (left panel) and 0.2-0.3 
(right panel) pulse phases. For the phase 0.2-0.3, we found that the 
(99\%) upper limits to N$_H$ and the blackbody normalization (N$_{BB}$) 
are found to be 3.8$\times$10$^{21}$ atoms cm$^{-2}$ and 2.4$\times$ 10$^{-5}$,
respectively. However, for phase 0.3-0.4, the acceptable ranges of N$_H$ and 
N$_{BB}$ are found to be (1.56-1.94) $\times$ 10$^{22}$ atoms cm$^{-2}$ and 
(4.5-9.5) $\times$ 10$^{-3}$, respectively (for 99\% confidence level). 
The confidence contours for above two phases show that the absorption column 
density and the blackbody flux are low at pulse phase 0.2-0.3 and high 
at 0.3-0.4 phase. During the dip-like structure, a hint of decrease in the 
blackbody flux and higher value of N$_H$ can be seen in the figure. This 
implies that the dip-like structure at soft X-rays in the pulse profile is 
due to the increase in the absorption column density resulting in the 
reduction of the apparent blackbody flux. The iron line equivalent width 
does not show any systematic variation over pulse phases. On the other hand, 
the modulation is a little larger in the line intensity, whose profile is 
similar to that of the NPEX flux. To statistically quantify these
variations, we estimated the values of $\chi^2$ by fitting a constant
model to the data. The results are 9.2 (for 9 degrees of freedom) for the
line intensity and 7.0 (9 degrees of freedom) for the equivalent width,
respectively. However, if we exclude the 0.2-0.3 phase bin, the $\chi^2$
value of the line intensity is reduced to 3.4 (8 degrees of freedom). If
we apply the F-test, the reduction corresponds to F$^{1}_{8}$=13.3 and 
probability
of 6.55$\times$10$^{-3}$. This means that the line intensity at phase 0.2-0.3
is significantly different from that of the other phases. In the case of
equivalent width, $\chi^2$ is 6.3 (8 degrees of freedom) even if we excluded
the 0.2-0.3 phase bin, which indicates no significant difference in the bin. 
The pulsating nature of the line flux suggests that the matter emitting the 
iron fluorescence line is distributed asymmetrically around the pulsar.

\section{Discussion}

Apart from two RXTE observations and three BeppoSAX observations during the 
quiescence, all other observations reported before 2005 August were 
made during the giant outbursts when the source flux was about equal to or 
more than that of the Crab nebula. The RXTE and INTEGRAL observations 
of the pulsar, during the 2005 August-September minor outburst, confirmed the 
earlier discovery of a cyclotron feature at $\sim$45 keV along with the 
detection of its first harmonic at $\sim$100 keV (Caballero et al. 2007). 
Suzaku observation of A0535+262 during the declining phase of the 2005 
August-September normal outburst provided new data on spectral properties 
of the pulsar up to $\sim$100 keV and how they might evolve throughout the 
outburst.

\subsection{Pulse Profile}

During the present observation of the minor outburst,
the pulse profile of A0535+262 is different from that in
the quiescence or during the earlier reported outbursts.
The pulse profile in the hard energy band is known to show
energy and luminosity dependence in outbursts, which is 
described in Bildsten et al. (1997) based on the BATSE 
observations. The profile tends to be double peaked when 
the luminosity is high, eg.\ $>\!10^{37}$ erg s$^{-1}$ 
(20--100 keV), whereas it becomes a single peak at 
$4\times10^{36}$ erg s$^{-1}$. The current $Suzaku$ observation 
had even lower luminosity ($2\times10^{35}$ erg s$^{-1}$ in 
20--100 keV band), and showed almost sinusoidal profile above 
25~keV\@. The $Suzaku$ hard band observation found a single
peaked profile, as was seen by BATSE.

Hard X-ray pulse profiles of A0535+262, obtained from the RXTE 
and INTEGRAL observations of the same 2005 outburst with a peak 
luminosity of $\sim$10$^{37}$ erg s$^{-1}$, appear to be double-peaked 
(Caballero et al. 2007). The two peaks with a shallow valley constitute
a trapezoid-like profile above 35 keV separated by a wide trough. Though 
the observations were carried out during the peak and the decay of the 
outburst, the added pulse profile may be dominated by the profile during 
the peak of the outburst. Therefore, the profile presented by Caballero 
et al. (2007) probably reflect that at the higher luminosity observations 
(at the peak of the outburst). The double-peaked hard X-ray profiles at a 
source luminosity of $\sim$10$^{37}$ erg s$^{-1}$ agrees with the luminosity 
dependence of pulse profiles observed with BATSE. The profiles obtained 
from Suzaku observation, however, appear to be different from that of the 
RXTE and INTEGRAL observations. This is interpreted due to the luminosity 
dependent effect as the Suzaku observation was at a luminosity level of 
two orders of magnitude less compared to the peak luminosity.

On the other hand, the Suzaku pulse profile in the soft energy band 
($<\!10$ keV) is found to be similar to that obtained from the 
$Ginga$ observation of the 1989 outburst (Mihara 1995). The dip-like 
structure, prominent only at soft X-rays is also recognized 
in the $Ginga$ profile. Because the $Ginga$ observation was made 
during the outburst when the source was $\sim$80 times brighter  
than the present observation, the profile may be interpreted that 
A0535+262 in outbursts generally contains a dip-like structure at 
soft X-rays as seen in 2005 $Suzaku$ observation. This type of
structure is also seen in the RXTE observation of the pulsar during
the same minor outburst in 2005 August/September (Caballero et al.
2007). This dip-like structure is absent in the pulse profiles when
the pulsar was in quiescence (Mukherjee \& Paul 2005). As other 
observations of A0535+262 in giant outbursts focused on the hard 
X-ray energy ranges, the dip-like structure could have been missed 
in the pulse profiles. 

\subsection{Phase-Averaged Spectroscopy}

The X-ray spectrum of A0535+262 
has been described by different continuum models at different energy 
ranges. Hard X-ray observations during outbursts, meant for the study 
of cyclotron absorption features and hence the pulsar magnetic field, 
were fitted by different continuum models such as optically thin thermal 
bremsstrahlung, power-law, power-law with an exponential cut-off,
Wien's law or different combinations of above components (dal Fiume et al.\
1988; Kendziorra et al.\ 1994; Grove et al.\ 1995), whereas the 2--37 keV
spectrum, obtained from the $Ginga$ observation of the 1989 outburst, was
well fitted with NPEX continuum model (Mihara 1995). Current analysis of 
the $Suzaku$ data showed that the NPEX continuum model fits well to the 
source spectrum. 

Selection of an appropriate continuum model may be crucial to investigate
the broad features in the spectrum, such as the soft excess represented by
a blackbody component. The spectral fitting to the 0.3--10~keV BeppoSAX 
spectra, during quiescence, yielded a blackbody component of temperature 
$\sim$1.3~keV along with a power-law component of index $\sim$1.8 (Mukherjee 
\& Paul 2005). On the other hand, the Suzaku data when fitted with NPEX 
continuum model yielded a much lower temperature of $\sim$0.16 keV, which 
is common for most of the X-ray pulsars. We suspect that the difference 
arises from the different selection of the continuum model. In fact, when 
the Suzaku wide band spectra were fitted by a power-law with an exponential 
cutoff, we did indeed obtain a blackbody temperature of 1.36 keV. This 
demonstrates the importance of the continuum selection to quantify the 
blackbody component. However, the selection criteria for the continuum
choice have yet to be rigorously defined.

Most of the transient Be/X-ray binary pulsars undergo periodic outbursts
due to the enhanced mass accretion when the neutron star passes through the
dense regions of the circumstellar disk. During the passage, there is every 
possibility of the increase in the absorption column density. In case of 
A0535+262, we found that the absorption column density was higher than the 
Galactic column density towards A0535+262 (5.9 $\times$ 10$^{21}$ atoms 
cm$^{-2}$) during the minor outburst in 2005 August-September. Note that
the $Suzaku$ observation was made near the apastron at the orbital phase
of 0.42--0.43. From BeppoSAX observations of the pulsar in quiescence 
(orbital phases 0.77, 0.05, 0.42), though, the estimated value of the 
absorption column density was found to be similar to that of the Galactic 
column density (Mukherjee \& Paul 2005). During the 1998 RXTE/PCA observations 
(in quiescence, orbital phases of 0.0 and 0.77), the value of the absorption 
column density was found to be a factor of about 10 higher than that of the 
Galactic column density (Negueruela et al. 2000). During the $Ginga$ 
observation of the pulsar in the 1989 outburst (orbital phase of 0.96), 
the estimated absorption column density was found to be 4.8$^{+1.7}_{-1.2}$
$\times$ 10$^{21}$ atoms cm$^{-2}$ (Mihara 1995), which is compatible with 
the Galactic value.  These are only three instances where the absorption 
column density of the pulsar is reported so far. Although the available
data are scarce, the absorption column density does not show clear correlation
with the source intensity or the orbital phase. In this sense, it is note
worthy that the observations of the large absorption column with RXTE/PCA
were made when the circumstellar disk around the Be companion (Be disk)
was absent (Negueruela et al. 2000). 

Haigh et al. (2004) suggest that the X-ray activity of the pulsar is 
correlated with the change of the truncation radius of the Be disk.  
The truncation of the disk by the tidal interaction with the neutron
star, which defines the truncation radius, could explain the observed 
X-ray outbursts in Be/X-ray binary systems. When the truncation radius 
reduces, the material from the outer portions of the disk are expelled 
and may be found elsewhere in the binary system. This may trigger the 
giant X-ray outburst. When the Be disk disappears, as the case of 
RXTE/PCA observations, all the disk material is considered to be 
accreted and/or ejected throughout the binary system. The 
presence of the expelled material in the line of sight may have 
caused a moderately large absorption column. Haigh et 
al. (2004) argues that the reduction of the Be disk continues for several 
orbital periods before/after the giant outburst. Because the Suzaku observation
was made about an orbital period after the giant outburst in May/June 2005,
the reduced Be disk may have caused moderately large absorption column.

Broad-band spectroscopy of A0535+262 shows the presence of a weak iron
K$_\alpha$ emission line with $\sim$25~eV equivalent width. Other than
the 1989 $Ginga$ observation and recent RXTE and INTEGRAL observations
of the same minor outburst, none of the previous observations of the
pulsar could detect the weak emission line at $\sim$6.4~keV\@.
The emission line is interpreted as due to the fluorescent line from
the cold ambient matter in the vicinity of the neutron star. If the cold 
matter is distributed spherically symmetric around the neutron star, the 
excess column density over the Galactic value, $\sim\!5\times10^{21}$ atoms 
cm$^{-2}$, would produce an iron emission line with an equivalent width of 
only a few eV (Makishima 1986). The larger equivalent width we detected may 
be interpreted as the cold matter having an asymmetric distribution, and 
more matter is distributed outside the line of sight. The fluorescence of 
the atmosphere of the binary companion could be a possible source of the 
iron emission line. However, we consider that it does not have a major 
contribution to the observed equivalent width, because the solid angle
subtended by the companion, which is estimated as $\Omega/4\pi$ $\leq$
10$^{-4}$ at apastron, is very small due to the long-orbital period of
A0535+262. Instead, we conjecture that the accretion disk is 
the probable reprocessing site to produce the iron fluorescent line. 
Because the pulse phase with higher line flux covers more than a half of
the pulse period, the reprocessing site should subtend a large solid angle
against the neutron star. The accretion disk conforms to this condition.

\subsection{Pulse Phase Resolved Spectroscopy}

Pulse phase resolved spectroscopy of A0535+262 showed that the
blackbody component, used to describe the soft excess, is pulsating as seen
in some other accreting X-ray pulsars, such as LMC~X-4, SMC~X-1 etc. (Naik
\& Paul 2004a, Naik \& Paul 2004b, Paul et al.\ 2002 and references
therein). The pulsation is found to be in phase with that of the middle-band
X-ray emission (eg. 2.5--5.0 keV band; see Figure~3). However, in case of 
Her~X-1, the pulsating soft component is found to be shifted by about 
230$^\circ$ in phase from the power-law continuum component (Endo et al. 
2000). A systematic and detailed study of the accreting X-ray pulsars which 
show soft excess revealed that the soft excess is a common feature in the 
X-ray pulsars (Hickox et al. 2004). Both the pulsating and non-pulsating 
natures of the soft excess seen in various X-ray pulsars suggest a different 
origin of emission of the soft components. The possible origins of the soft 
excess in accretion powered X-ray pulsars are (a) emission from accretion 
column, (b) emission by a collisionally energized cloud, (c) reprocessing 
by a diffuse cloud, and (d) reprocessing by optically thick material in 
the accretion disk (Hickox et al. 2004). The soft excess emission from the 
accretion column and by the reprocessing of harder X-rays by optically thick 
material in the accretion disk, are expected to show pulsations whereas in 
other cases, it is non-pulsating in nature. The pulsation of the soft 
component (blackbody component) in phase with the middle-band X-ray emission 
in A0535+262 suggests that the most probable region of soft X-ray emission is 
the accretion column and/or the inner accretion disk.

The absorption column seems to have significant contribution to
produce the dip like structure at soft X-rays at phases 0.65-0.80
in the count rate profile (left top panel of Figure~\ref{phrs}). Column 
density takes large values, $\sim\!1.2\times10^{22}$ atoms cm$^{-2}$, 
at phase 0.7-0.8. This may contribute to produce the dip like structure
in the profiles below 1 keV which disappears from the profiles at higher
energies. On the other hand, the primary dip exists at all the energies. 
The observed primary dip in the pulse profile can be understood by 
the change in the NPEX model parameters over pulse phase. The NPEX continuum 
model is an approximation of the unsaturated thermal Comptonization in hot 
plasma (Makishima et al. 1999). At low energies, it reduces to the ordinary 
power-law with negative slope. The lower value of the power-law index 
($\alpha_1$) at the primary dip phase implies large optical depth 
to the Compton scattering. This means that many photons are scattered-out
from the line of sight. It explains the presence of the primary dip
in the soft and hard X-ray pulse profiles at same phase. In addition to
the primary dip, the deepening of the dip at the phases 0.0-0.17 
(Figure~\ref{erpp}) can also be understood by the changes of the 
parameters of the Comptonizing plasma. The power-law index ($\alpha_1$) 
tends to be larger and the exponential cut-off energy is smaller in this 
phase range. This means that the Comptonizing plasma has a relatively 
smaller optical depth and a lower temperature. This reduces the efficiency 
of Compton up-scattering and the number of hard photons, which causes the 
deepening of the dip at phases 0.0-0.17 above 5 keV.

\section{Summary}
Using the Suzaku observation of A0535+262, we performed the timing and spectral
analysis in broad band energy range. The $\sim$103 s pulsation is detected both
in HXD/PIN and XIS light curves of the pulsar. Apart from the primary dip, 
the pulse profile at soft X-rays shows a dip like structure which disappears 
at higher energies. When the source spectrum was modeled by a cut-off
power-law continuum model with a blackbody component for the soft excess, it 
yielded a blackbody temperature of $\sim$1.36 keV. However, the NPEX 
continuum model provided a blackbody component of $\sim$0.16 keV temperature,
which is common in case of accretion powered X-ray pulsars, along with a weak 
iron emission line of equivalent width of about 25 eV. The value of the 
absorption column density (N$_H$) is found to be higher than that of the 
galactic value towards the pulsar. The pulsating nature of the blackbody 
component, as seen from the phase resolved spectroscopy, suggests that the 
accretion column and/or inner part of the accretion disk is the possible 
source of soft excess emission in A0535+262. The iron line equivalent width 
remains more or less constant over the pulse phase with an average value of 
about 25 eV. This value of equivalent width is found to be high for a 
symmetric distribution of circumstellar matter corresponding to N$_H$ of 
5$\times$10$^{21}$ atoms cm$^{-2}$. This rules out the fluorescence from 
the atmosphere of the companion and fluorescence by the surrounding material
in the line of sight as the dominant source of iron line emission. 
The pulsating nature of the iron line flux in phase with the hard X-ray 
(NPEX) flux suggests that the inner part of the accretion disk 
is the probable site of iron fluorescence emission.


\section*{Acknowledgments}
The authors would like to thank all the members of the $Suzaku$ Science 
Working Group for their contributions in the instrument preparation, 
spacecraft operation, software development, and in-orbit instrumental 
calibration. The authors also thank the referee for useful comments and 
suggestions. SN acknowledges the support by JSPS (Japan Society for 
the Promotion of Science) post doctoral fellowship for foreign researchers
(P05249). This work was partially supported by grant-in-aid for JSPS fellows
(1705249).

\end{document}